\documentclass{iopart}

\usepackage[latin1]{inputenc}
\usepackage[english]{babel}
\usepackage{graphicx}
\usepackage{setstack}
\usepackage{amsopn}
\usepackage{amstext}
\usepackage{amssymb}
\usepackage{amsbsy}
\usepackage{float}
\usepackage{mathrsfs}
\usepackage{bbm}
\usepackage{color}
\usepackage{caption,subfig}

\def\ket#1{\vert #1\rangle}
\def\bra#1{\langle #1\vert}
\def\ketbra#1#2{\vert #1\rangle \langle #2\vert}

\DeclareMathOperator{\Var}{Var}

\newcommand{\fiftyfifty}{$50$\,:\,$50$ }

\begin{document}

\title[Gaussian Entanglement for Quantum Key Distribution]{Gaussian Entanglement for Quantum Key Distribution from a Single-Mode Squeezing Source}

\author{Tobias Eberle$^{1,2}$, Vitus H\"andchen$^1$, J\"org Duhme$^{2,3}$, \\Torsten Franz$^3$, Fabian Furrer$^4$, Roman Schnabel$^1$, Reinhard F Werner$^3$}
\address{$^1$ Max-Planck-Institut f\"ur Gravitationsphysik
(Albert-Einstein-Institut) and\\ Institut f\"ur Gravitationsphysik,
Leibniz Universit\"at Hannover, Callinstra{\ss}e~38, 30167 Hannover, Germany}
\address{$^2$ Centre for Quantum Engineering and
Space-Time Research - QUEST, Leibniz Universit\"at Hannover,
Welfengarten 1, 30167 Hannover, Germany}
\address{$^3$ Institut f\"ur Theoretische Physik, Leibniz Universit\"at Hannover, Appelstra{\ss}e 2, 30167 Hannover, Germany}
\address{$^4$ Department of Physics, Graduate School of Science, University of Tokyo, 7-3-1 Hongo, Bunkyo-ku, Tokyo, Japan, 113-0033}
\ead{roman.schnabel@aei.mpg.de}

\begin{abstract}
    We report the suitability of an Einstein-Podolsky-Rosen (EPR) entanglement source for gaussian continuous-variable quantum key distribution at $1550$\,nm. Our source is based on a single continuous-wave squeezed vacuum mode combined with a vacuum mode at a balanced beam splitter. Extending a recent security proof, we characterize the source by quantifying the extractable length of a composable secure key from a finite number of samples under the assumption of collective attacks. We show that distances in the order of $10$\,km are achievable with this source for a reasonable sample size despite the fact that the entanglement was generated including a vacuum mode. Our security analysis applies to all states having an asymmetry in the field quadrature variances, including those generated by superposition of two squeezed modes with different squeezing strengths.
\end{abstract}


\maketitle

\section{Introduction}
Quantum key distribution (QKD) enables two remote parties to generate a shared key which is guaranteed to be unknown to any potential eavesdropper. Discrete variable systems implementing, for example, the famous \emph{BB84} protocol~\cite{BB84} are well established~\cite{Gisin2002} and commercial systems exist. Recently, also first commercial continuous variable (CV) systems have been launched in which the field quadratures of laser light are measured by homodyne detection. Compared to discrete variable systems, they have the advantage that for homodyne detection PIN photo diodes can be used which are well developed and widely used telecommunication components. They offer high bandwidth, low dark noise and high quantum efficiencies. Most of todays CV quantum key distribution systems use prepare-and-measure schemes employing coherent states with gaussian or discrete modulation~\cite{Grosshans2002}\nocite{Lodewyck2007,Fossier2009,Lance2005}\,-\,\cite{Leverrier2009}. Prepare-and-measure schemes with squeezed states have been considered in~\cite{Hillery2000,Gottesman2001}. The less common entanglement-based schemes do not need signal modulation \cite{Rodo2007}, and instead exploit directly the correlations in the field quadratures of an Einstein-Podolsky-Rosen (EPR) entangled state. EPR entangled states are usually generated by interfering two squeezed beams at a beam splitter~\cite{Ou1992}\,-\,\cite{Silberhorn2001}.

An implementation of a suitable source for an CV entanglement-based scheme was shown in~\cite{DiGuglielmo2007} and a demonstration of a fully implemented table-top QKD was done in~\cite{Su2009}. In both cases the security analysis assumed an infinite number of measured samples which is experimentally unfeasible. Security proofs for CV systems including the effect of a finite number of samples were only recently published~\cite{Leverrier2010,Furrer2012}. In~\cite{Jouguet2012} an experiment including finite-size effects was performed using gaussian modulation and coherent states with a security analysis under the assumption of collective attacks.

In this paper we characterize EPR entangled states generated by superimposing a squeezed vacuum mode with a vacuum mode at a balanced beam splitter in terms of extractable key length. Using only one squeezed mode instead of two minimizes the necessary resources and reduces the complexity of the setup. Our source is implemented at the telecommunication wavelength of $1550$\,nm to, in principle, allow for efficient coupling to existing telecommunication fiber networks. We calculate the extractable key rate as a function of measured samples for various communication distances through an optical fiber. We assume that the source is located in the lab of an honest party such that only one part of the beam is affected by transmission losses. The key rate is computed by applying the security proof for composable security under the assumption of collective attacks including finite-size effects given in~\cite{Furrer2012} to states with asymmetric field quadrature variances. As asymmetries in the field quadrature variances are experimentally unavoidable even for entanglement generated by two squeezed modes, our analysis can also be applied to such states.

The paper is organized as follows: In Section~\ref{sec: security analysis} we describe the protocol, give the security definitions and extend the proof given in~\cite{Furrer2012} to asymmetric states. Section~\ref{sec: experiment} is devoted to the details of the experimental setup. The main results are presented in Section~\ref{sec: results}. The paper ends with the conclusions in Section~\ref{sec: conclusion}.

\section{Security Analysis}
\label{sec: security analysis}
In this section we extend the security proof for composable security against collective Gaussian attacks including finite-size effects given in~\cite{Furrer2012} towards two-mode squeezed states with an asymmetry of the noise distributions of the field quadratures. By superimposing a squeezed mode with a vacuum mode at a balanced beam splitter the two output modes are still squeezed in one quadrature which we assume to be the amplitude quadrature $X$, and anti-squeezed in the orthogonal quadrature, the phase quadrature $P$.

\subsection{Protocol}
The protocol we use goes as follows:
\begin{enumerate}
    \item Preparation and measurement: Alice prepares an entangled state with her EPR source, keeps one subsystem and sends the other to Bob. Both parties perform homodyne measurements in either the $X$ or $P$ quadrature which is individually chosen at random. An outcome of such a synchronous measurement is called a sample. This process is repeated until $2N$ samples were recorded, forming two strings $x_A^{\prime\prime}$ and $x_B^{\prime\prime}$ of length $2N$.
    \item Sifting: In the second step Alice and Bob perform sifting, i.e.\ they communicate which quadrature they measured. Samples measured with a different choice of quadrature are discarded from $x_A^{\prime\prime}$ and $x_B^{\prime\prime}$ leaving Alice and Bob with strings $x_A^\prime$ and $x_B^\prime$ of length $N$ in average. The discarded data is used for parameter estimation.
    \item Parameter estimation: In the third step Alice and Bob choose randomly a common subset of length $k$ from $x_A^\prime$ and $x_B^\prime$ which they reveal. From this data and the data discarded by the sifting procedure, they reconstruct the covariance matrix. In particular, they estimate a confidence set $ \mathcal{C}_{\epsilon_{\text{pe}}}$ with the property that with probability $1-\epsilon_{\text{pe}}$ the real covariance matrix lies within $ \mathcal{C}_{\epsilon_{\text{pe}}}$.
    \item (Optional) Discarding $X$ or $P$ quadrature measurements: As the $X$ and $P$ quadrature measurements of Alice and Bob are correlated with different strength due to the asymmetric nature of the bipartite entangled state, it might be beneficial to discard the measurements performed in the $P$ quadrature from the raw key. To take into account all three possibilities, i.e.\ discarding $X$ measurements, discarding $P$ measurements and discarding nothing at all, we introduce a parameter $p_X$ which describes the probability of a sample being measured in the $X$ quadrature. For taking both $X$ and $P$ quadrature measurements to generate a raw key $p_X \approx 0.5$ depending on the actual run. By discarding all $X$ measurements $p_X = 0$ and $p_X=1$ for dicarding all $P$ quadrature measurements. The number of samples left after this step is denoted by $n$.
    \item Binning: In the fourth step Alice and Bob group their unrevealed samples into bins
    $\left(-\infty,-\alpha_{\{X,P\}}+\delta_{\{X,P\}}\right)$, $\left(-\alpha_{\left\{X,P\right\}}+\delta_{\left\{X,P\right\}},-\alpha_{\left\{X,P\right\}}+2\delta_{\left\{X,P\right\}}\right)$, $\dots$, $\left(\alpha_{\left\{X,P\right\}}-\delta_{\left\{X,P\right\}}, \infty\right)$. Each bin is assigned a unique bit combination so that after the conversion Alice and Bob both have a bit string representing their raw key. In practice, we always choose $\alpha$ such that for no sample the quadrature measurement exceeded $\alpha$. In that sense, only $\delta$ is a free parameter in the protocol.

\item Classical Postprocessing: In the last step Alice and Bob perform error correction and privacy amplification to extract $\ell$ secure bits. We assume that they execute a reverse information reconciliation protocol (error correction) in which Bob only sends information to Alice and denote the number of bits revealed by Bob by $\ell_{\text{EC}}$. In the final privacy amplification step both parties apply two-universal hash functions to reduce the key length to $\ell$ bits, where $\ell$ is computed according to Eq.~(\ref{eqn:key rate}).
\end{enumerate}

\subsection{Security definitions}

It is important to ensure that the key is secure in any further cryptographic sub-protocol like for instance the one-time pad to securely transmit messages between Alice and Bob. To guarantee this we use the composable security definitions from~\cite{Canetti2001,Renner2005}. In the following, we denote by $S_A$ and $S_B$ the random variables associated with the final key of Alice and Bob at the very end of the protocol.

\paragraph{Robustness.} We call a protocol robust if it does not abort when no eavesdropper is present. This ensures that the protocol is not trivial.
\paragraph{Correctness.} A protocol is $\epsilon_c$-correct if
\begin{equation*}
    \text{Prob}[S_A \neq S_B] \le \epsilon_c\ .
\end{equation*}
If $\epsilon_c\ll 1$, this implies that Alice's and Bob's key agree with high probability.
\paragraph{Secrecy.} Let $\omega_{S_BE}$ denote the classical-quantum state of Bob's final key $S_B$ and a possible eavesdropper $E$. Such a state can always be written as
\begin{equation*}
    \omega_{S_BE} = \sum_{s_B \in S_B}p(s_B)\ket{s_B}\bra{s_B} \otimes \omega_E^{s_B}\ ,
\end{equation*}
where $p(s_B)$ is the probability distribution of the key. We then call a protocol $\epsilon_s$-secret if for any eavesdropper $E$
\begin{equation*}
    \frac{p_{\text{pass}}}{2}\Vert\omega_{S_BE} - \tau_{S_B} \otimes \omega_E\Vert_1 \le \epsilon_s\
\end{equation*}
holds. Here, $\Vert\cdot \Vert_1$ is the trace norm, $\tau_{S_B}$ is the uniform distribution over $S_B$, $\omega_E$ is the reduced state of $\omega_{S_BE}$ and $1-p_{\text{ pass}}$ is the probability that the protocol aborts.

\paragraph{Security} A protocol is $\epsilon$-secure if it is $\epsilon_c$-correct and $\epsilon_s$-secret with $\epsilon_c + \epsilon_s \le \epsilon$.

For a detailed discussion of the above security conditions we refer to~\cite{MuellerQuade2009}.

\subsection{Secure Key Rates}
Let us consider the stage of the protocol before applying error correction and privacy amplification. We call the remaining $n$ samples at this stage the raw keys and denote the corresponding random variables on Alice's and Bob's side by $X^n_A$ and $X^n_B$. One can assume that $X^n_A$ and $X^n_B$ are obtained by performing $n$ times a quadrature measurement where amplitude $X$ is chosen with probability $p_X$ and phase $P$ with $p_P=1-p_X$. Note that for the following security discussion $p_X$ can be arbitrarily chosen. Let in the following $E^n$ be the eavesdropper system which can be infinite-dimensional and $\omega_{X^n_AX^n_BE^n}$ the corresponding classical-quantum state conditioned on the event that the protocol passes. It was shown in~\cite{Berta2011} that an $\epsilon_c$-correct and $\epsilon_s$-secret key of length
\begin{equation}
    \max_{\epsilon_1} \left[H_{\text{min}}^\epsilon(X_B|E)_{\omega} - \ell_{\text{EC}} - \log_2 \frac{1}{4\epsilon_1^2\epsilon_c}\right]
\end{equation}
can be extracted. $H_{\text{min}}^\epsilon(X_B^n|E^n)_{\omega}$ denotes the conditional smooth min-entropy of $\omega_{X_B^nE^n}$ for $\epsilon \le (\epsilon_s-\epsilon_1)/2$ introduced in~\cite{Renner2005} and generalized to infinite-dimensional systems in~\cite{Berta2011,Furrer2011}. Hence, it remains to obtain a lower bound on $H_{\text{min}}^\epsilon(X_B^n|E^n)_{\omega}$ for any possible eavesdropping strategy.

Under the assumption of collective attacks, we can assume that the state $\omega_{X_A^nX_B^nE^n}$ has tensor product structur, i.e., $\omega_{X_A^nX^n_BE^n}=\omega_{X_AX_BE}^{\otimes n}$. The smooth min-entropy of a product state can then be approximated by the conditional von Neumann entropy $H(X_B|E)_\omega$ of $\omega_{X_BE}$ via the asymptotic equipartition property~\cite{Furrer2011}
\begin{equation}
    \label{eqn:generic key rate}
    H_{\text{min}}^\epsilon(X_B|E)_{\omega} \le n  H(X_B|E)_{\omega} - \sqrt{n}\Delta\ ,
\end{equation}
where $n$ has to be sufficiently large and
\begin{equation*}
    \Delta = 4\log_2\left(2^{\frac{1}{2}H_{\text{max}}(X_B)+1}+1\right)\sqrt{\log_2\frac{2}{\epsilon^2}}\ .
\end{equation*}
In the next step, we use that the state $\omega_{X_BE}$ is of form $\omega_{X_BE} =  p_X\ketbra XX_\theta\otimes \omega_{X_BE}^X + (1-p_X)\ketbra PP_\theta \otimes \omega_{X_BE}^P$ where $\omega_{X_BE}^X$, $\omega_{X_BE}^P$ are the states obtained when the honest parties are measuring amplitude or phase, respectively. The system denoted by $\theta$ is a classical register which is assigned to the eavesdropper and keeps track which measurements were performed by the honest parties. Using elementary properties of the von Neumann entropy, we can now expand $H(X_B|E\theta)_{\omega} = p_X H(X_B|E)_{\omega^X} + p_P H(X_B|E)_{\omega^P}$.
Combining this estimation of the smooth min-entropy with the assumption of Gaussian attacks, we can use the confidence set $\mathcal{C}_{\epsilon_{\text{pe}}}$ to obtain a lower bound on the key length given by
\begin{equation}
    \label{eqn:key rate}
     \ell = n\cdot \inf_{\gamma \in \mathcal{C}_{\epsilon_{\text{pe}}}} \sum_\theta p_\theta H(X_B|E)_{\omega^{\gamma,\theta}} - \sqrt{n}\Delta - \ell_{\text{EC}} - \log_2\frac{1}{\epsilon_s^2\epsilon_c} \ .
\end{equation}
Here, the infimum is taken over all states compatible with covariance matrices $\gamma$ within the confident set. For simplicity, we have chosen $\epsilon_1= \epsilon_s/2$ which can be justified by the fact that for large enough $n$ the term in the logarithm can be neglected. Note further that due to the definition of $\mathcal{C}_{\epsilon_{\text{pe}}}$, the key length from Eq.~(\ref{eqn:key rate}) is now $\epsilon$-secure with $\epsilon = \epsilon_{\text{pe}} + \epsilon_s + \epsilon_c$.

The von Neumann entropy for both quadratures $\theta = X,P$ can now be computed under the non-restricting assumption that the eavesdropper holds the purification of Alice's and Bob's state, that is, we assume that $\omega^\gamma_{ABE}$ is the purification of the Gaussian state $\omega_{AB}^\gamma$ with covariance matrix $\gamma$. It then follows by applying the definition of the conditional von Neumann entropy $H(X_B|E) = H(X_BE) - H(E)$ and the self-duality $H(E)_{\omega^\gamma}=H(AB)_{\omega^\gamma}$ that
\begin{equation}
H(X_B|E)_{\omega^{\gamma,\theta}} = H(E|X_B)_{\omega^{\gamma,\theta}} + H(X_B)_{\omega^{\gamma,\theta}} - H(AB)_{\omega^\gamma} \, .
\end{equation}
As shown in~\cite{Furrer2012,Eisert2002,Fiurasek2002}
\begin{eqnarray*}
    H(E|X_B)_{\omega^{\gamma,X}} \ge H(E)_{\omega^{\gamma,X}(X=0)} = H\left(A-C(M_XBM_X)^{\text{MP}}C^T\right)_{\omega^{\gamma}}
\end{eqnarray*}
and
\begin{equation*}
    H(E|X_B)_{\omega^{\gamma,P}} \ge H(E)_{\omega^{\gamma,P}(P=0)} = H\left(A-C(M_PBM_P)^{\text{MP}}C^T\right)_{\omega^\gamma}\ ,
\end{equation*}
where $H(E)_{\omega^{\gamma,\{X,P\}}(\{X,P\}=0)}$ is the post-measurement state at the eavesdropper's side when Bob measured $X=0$ or $P=0$. The bipartite covariance matrix is written in block form, i.e.
\begin{equation*}
    \gamma = \left(\begin{array}{cc} A & C\\C^T&B\end{array}\right)
\end{equation*}
and $M_X = \text{diag}(1,0)$ and $M_P = \text{diag}(0,1)$ are the projectors to the $X$ and $P$ quadrature, respectively. $\text{MP}$ denotes the Moore-Penrose inverse.

To compute $\Delta$, we have to estimate $H_{\text{max}}(X_B)$ which can be approximated by~\cite{Furrer2011}
\begin{eqnarray*}
    H_{\text{max}}(X_B) \le &2\log_2\Big(\sqrt{p_X}\sum_{y}\sqrt{\omega^X_{X_B}(y)} + \sqrt{(1-p_X)}\sum_{y}\sqrt{\omega^P_{X_B}(y)}\Big)\ ,
\end{eqnarray*}
where $\omega^X_{X_B}$ and $\omega^P_{X_B}$ are the probability distributions of Bob's $X$ and $P$ quadrature measurements, respectively.

While in a practical experiment the number of bits $\ell_{\text{EC}}$ can be directly measured for each run, we need to estimate the leakage term here. We assume the term to be~\cite{Scarani2009}
\begin{eqnarray*}
    \ell_{\text{EC}} = &p_X\left(H(X_B)_{\omega^{\gamma,X}} - \beta I(X_A,X_B)_{\omega^{\gamma,X}}\right)\\
    &+ (1-p_X)\left(H(X_B)_{\omega^{\gamma,P}} - \beta I(X_A,X_B)_{\omega^{\gamma,P}}\right)\ ,
\end{eqnarray*}
where $\beta \in (0,1)$ is the error correction efficiency and $I(A,B)$ is the mutual information. In this paper we will assume an error correction efficiency of $\beta=0.9$~\cite{Martinez-Mateo2012}.

With these results the secure key rate $r = \ell/n$ can be calculated by
\begin{eqnarray*}
    r = &\inf_{\gamma\in \mathcal{C}_{\epsilon_{\text{pe}}}}p_X\left[H(E|X_B)_{\omega^{\gamma,X}} + H(X_B)_{\omega^{\gamma,X}}\right]\\
    &+ (1-p_X)\left[H(E|X_B)_{\omega^{\gamma,P}} + H(X_B)_{\omega^{\gamma,P}}\right]\\
    &- H(AB)_{\omega^\gamma} - \frac{1}{\sqrt{n}}\Delta - \frac{\ell_{\text{EC}}}{n} - \frac{1}{n}\log_2\frac{1}{\epsilon_s^2\epsilon_c}
\end{eqnarray*}
In the theoretical asymptotic limit for an infinite number of samples $n \rightarrow \infty$ and perfect security $\epsilon\rightarrow 0$, the key rate $r$ tends to
\begin{eqnarray*}
    r_{\infty} = &p_X \left( \beta I(X_A,X_B)_{\omega^{\gamma,X}} + H(E)_{\omega^{\gamma,X(X=0)}} - H(AB)_{\omega^\gamma}\right)\\
     &+ (1-p_X)\left(\beta I(X_A,X_B)_{\omega^{\gamma,P}} + H(E)_{\omega^{\gamma,P(P=0)}} - H(AB)_{\omega^\gamma}\right)\ .
\end{eqnarray*}
Note that in the asymptotic limit the error correction protocol achieves the Shannen rate, and thus, one could set $\beta = 1$. However, since we only use the asymptotic limit to compare the key rate to the finite-size effects which are not connected to the efficiency of the error correction protocol, we treat $\beta$ as a constant over all sample sizes.

\subsection{Parameter Estimation}
To calculate the secure key rate we need to construct the confidence set $\mathcal{C}_{\epsilon_{\text{pe}}}$ which is defined such that the covariance matrix describing the real state lies within $\mathcal{C}_{\epsilon_{\text{pe}}}$ with probability $1-{\epsilon_\text{pe}}$. As our states are two-mode squeezed vacuum states, the first moment vanishes and the state is fully described by its covariance matrix. It is reconstructed during the parameter estimation step from the discarded samples and the revealed common subset of length $k$ using a maximum likelihood estimator. The sample covariance matrix is then estimated by
\begin{equation*}
    \tilde\gamma_{\mu\nu} = \frac{1}{n_{\mu\nu}}\sum_{i=1}^{n_{\mu\nu}} x_i^\mu x_i^\nu\ ,
\end{equation*}
where $x_i^\mu$ and $x_i^\nu$ are the samples measured simultaneously by Alice and Bob in $\mu$ and $\nu$ quadrature, respectively. $n_{\mu\nu}$ is the number of samples used for the covariance estimation which might in our case be different for different entries. The distribution of the sample covariance matrix $\tilde\gamma$ is given by~\cite{Johnson2007}
\begin{equation*}
  n\tilde\gamma \sim W_4(\gamma,n-1)\ ,
\end{equation*}
where $W_4(\gamma,n-1)$ is the Wishart distribution. Hence, the standard deviation for a single entry of the covariance matrix takes the form
\begin{equation*}
    \sigma_{\mu\nu} \approx \sqrt{\frac{\tilde\gamma_{\mu\nu}^2+\tilde\gamma_{\mu\mu}\tilde\gamma_{\nu\nu}}{n_{\mu\nu}}}\ .
\end{equation*}
For a large enough number of samples the confidence set is then constructed by
\begin{equation}
    \mathcal{C}_{\epsilon_{\text{pe}}} = \left\{\gamma | \tilde\gamma_{\mu\nu}-z_{\epsilon_{\text{pe}}}\sigma_{\mu\nu} \le \gamma_{\mu\nu} \le \tilde\gamma_{\mu\nu}+z_{\epsilon_{\text{pe}}}\sigma_{\mu\nu}\right\}\ ,
\end{equation}
where $z_{\epsilon_{\text{pe}}}$ is chosen such that
\begin{equation*}
    1-\text{erf}\left(\frac{z_{\epsilon_{\text{pe}}}}{\sqrt{2}}\right) \le \epsilon_{\text{pe}}\
\end{equation*}
is fulfilled. Here, $\text{erf}$ is the error function which is defined by
\begin{equation*}
    \text{erf}(x) = \frac{2}{\sqrt{\pi}}\int_0^x\text{d} t \exp\left(-t^2\right)\ .
\end{equation*}

\section{Experiment}
\label{sec: experiment}

\begin{figure}[ht]
  \center
  \includegraphics[width=13cm]{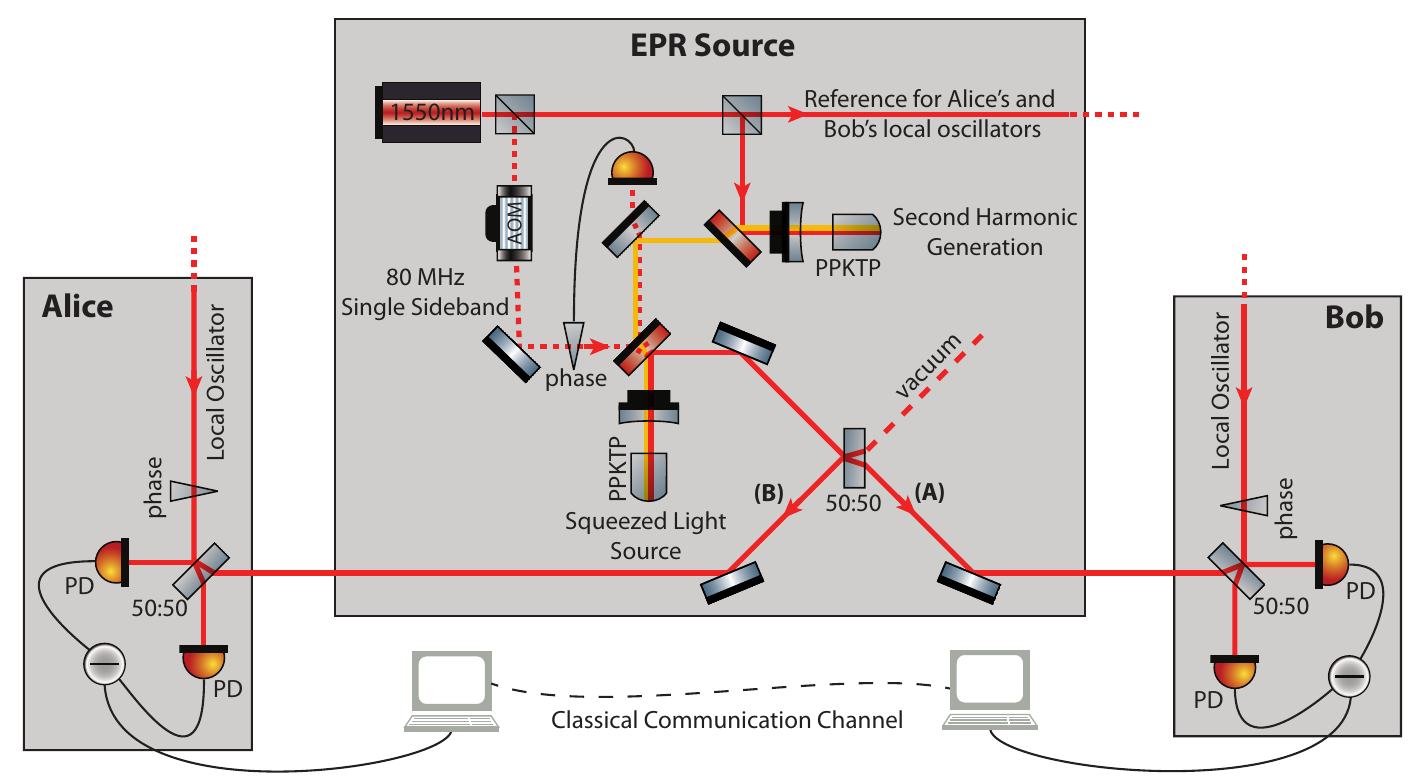}
  \caption{Schematic of the experiment: The beam of a $1550$\,nm fiber laser (red) was frequency doubled (yellow) and used as a pump for the squeezed-light source. The squeezed beam was overlapped with a vacuum mode at a \fiftyfifty beam splitter to produce a pair of EPR entangled beams. The field quadratures of both beams were measured by balanced homodyne detection to characterize the EPR source and to provide data points that can be used to extract a secret quantum key from simultaneous measurements of the amplitude or phase quadrature. AOM: acousto-optical modulator, PD: photo diode.}
  \label{fig:experiment}
\end{figure}

A schematic of the experiment is shown in Fig.~\ref{fig:experiment}. The EPR entanglement source was driven by a commercial $1$\,W $1550$\,nm fiber laser. Most of its power was frequency doubled in a quasi phase-matched periodically poled potassium titanyl phosphate (PPKTP) crystal~\cite{Ast2011} and served as a pump for the squeezed-light source which consisted of a $1 \times 2 \times 9.3$\,mm$^3$ PPKTP crystal. One of the end faces of the squeezed-light source's crystal was curved with a radius of curvature of $12$\,mm and coated with a high-reflective coating for both the pump and the fundamental beam at $775$\,nm and $1550$\,nm, respectively. The other end face was flat and anti-reflective coated for both wavelengths. Together with a coupling mirror with a radius of curvature of $25$\,mm a hemilithic cavity was formed. The coupling mirror had a reflectivity of $90$\,$\%$ for $1550$\,nm and a reflectivity of $20$\,$\%$ for $775$\,nm. With a $23$\,mm air gap between the crystal and the coupling mirror the cavity had a finesse of $60$ at $1550$\,nm, a free spectral range of $3.8$\,GHz and a full width half maximum linewidth of $63$\,MHz. The temperature of the PPKTP crystal was tuned to about $50^\circ$\,C to achieve quasi phase-matching. A sub-milliwatt control beam which was coupled into the cavity through the high-reflective mirror, was used to lock both the length of the cavity and the phase of the pump. The output of the squeezed-light source was split from the pump by a dichroic beam splitter and superimposed with a vacuum mode at a balanced beam splitter to produce a pair of EPR entangled beams. The field quadratures of these beams were measured by homodyne detection. For this each beam was overlapped with a strong local oscillator of about $10$\,mW at a balanced beam splitter with a visibility of about $99.5$\,$\%$ and detected by a pair of custom-made PIN photo diodes with high quantum efficiency. By changing the relative phase between the local oscillator and the quantum field the measured field quadrature could be chosen. Whereas for quantum key distribution Alice and Bob only have to randomly measure the amplitude and phase quadrature, also a linear combination of these is needed to reconstruct the full covariance matrix. Therefore we implemented a single sideband technique to be able to lock both homodyne detectors independently to any quadrature angle. An $80$\,MHz frequency shifted beam, produced by an acousto-optical modulator, was coupled into the squeezing path through the dichroic beam splitter and was phase locked to the control beam leaking through it. The single sideband was detected by the homodyne detectors and demodulated at $80$\,MHz. By choosing the phase of the electronic oscillator used for the demodulation the homodyne detector could be set to measure any field quadrature angle.

The outputs of both homodyne detectors were recorded simultaneously by a data acquisition system for which they were demodulated with a double-balanced mixer at $8.3$\,MHz and lowpass filtered with an anti-aliasing filter with a passband of $40$\,kHz. The data was sampled with $14$\,bit resolution at a sampling rate of $500$\,kHz.

\section{Experimental Results}
\label{sec: results}
\begin{figure}[p]
  \center
  \subfloat[Includes only $X$ quadrature measurements, i.e.\ $p_X=1$.]{\includegraphics[width=8.5cm]{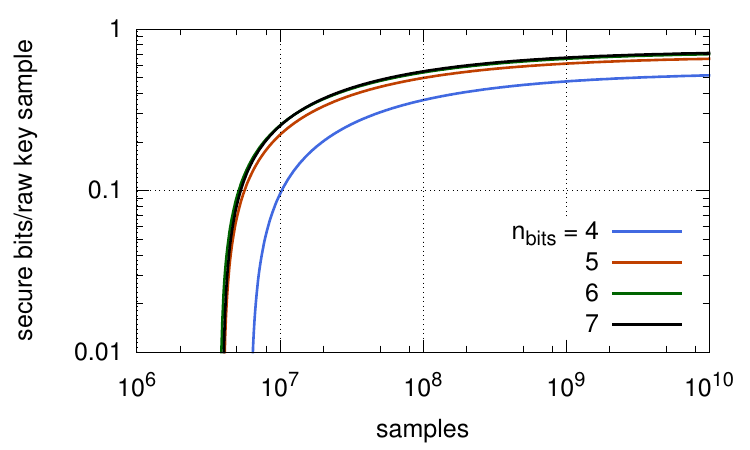}\label{fig:keyrate_vs_samples_X}}\\
  \subfloat[Includes only $P$ quadrature measurements, i.e.\ $p_X = 0$.]{\includegraphics[width=8.5cm]{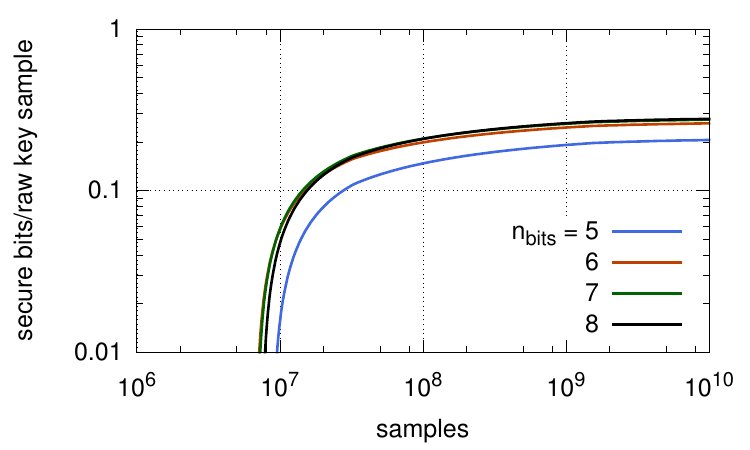}\label{fig:keyrate_vs_samples_P}}\\
  \subfloat[Includes both $X$ and $P$ quadrature measurements, i.e.\ $p_X = 0.5$. $n_{\text{bits}}=6$ for the $X$ quadrature and $n_{\text{bits}}=7$ for the $P$ quadrature.]{\includegraphics[width=8.5cm]{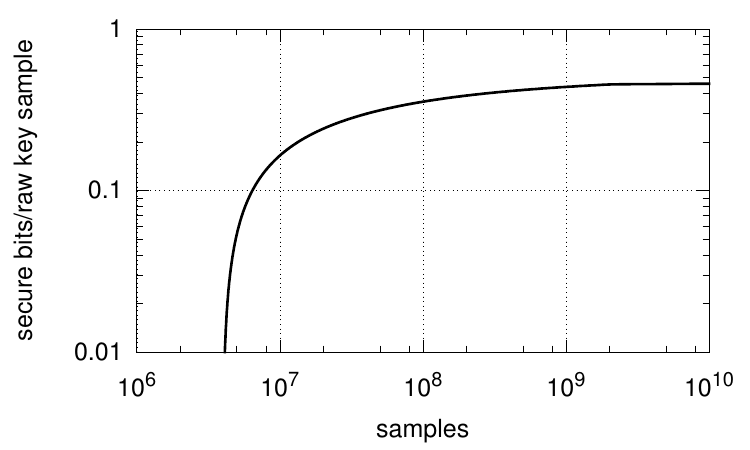}\label{fig:keyrate_vs_samples_XP}}
  \caption{Secure key rate versus the number of measured samples. The key rate on the y-axis is given in secure bits per raw key sample, i.e.\ it is normalized to the number of samples left after sifting and parameter estimation.}
  \label{fig:keyrate_vs_samples}
\end{figure}
To characterize our EPR source we fully reconstructed the covariance matrix of the bipartite state. The full reconstruction of the covariance matrix followed a protocol described in~\cite{DiGuglielmo2007}, where the complete covariance matrix of a gaussian state was measured for the first time. The protocol goes as follows:
\begin{enumerate}
  \item Alice and Bob both measure the amplitude quadrature.
  \item Alice and Bob both measure the phase quadrature.
  \item Alice measures the amplitude quadrature, whereas Bob simultaneously measures the phase quadrature.
  \item Alice measures the phase quadrature, whereas Bob simultaneously measures the amplitude quadrature.
  \item Alice and Bob both measure a linear combination of the amplitude and phase quadrature. In our case we chose the $45^\circ$ angle for both parties.
\end{enumerate}

Following the protocol above we recorded $5\times 10^6$ samples for each quadrature combination using a pump power of $235$\,mW for the squeezed-light source which allowed the observation of $11.1$\,dB squeezing and $16.6$\,dB anti-squeezing. From these data we reconstructed the covariance matrix which reads
\begin{equation}
    \label{eqn:covariance matrix}
  \Gamma =
  \left(\begin{array}{cc|cc}
  0.541 & 0.135 & 0.459 & -0.095\\
  0.135 & 24.633 & -0.037 & -23.293\\\hline
  0.459 & -0.037 & 0.548 & 0.264\\
  -0.095 & -23.293 & 0.264 & 23.840\\
  \end{array}\right)\ .
\end{equation}
One can directly see certain properties of the state from the entries in the matrix. The values on the principal diagonal are the variances for the amplitude and phase quadrature measurements at Alice's and Bob's detector. The diagonal entries of the two blocks in the upper right and lower left give the strengths of the correlations in the amplitude quadrature and the anti-correlations in the phase quadrature, respectively, between both detectors. In a perfect orthogonal measurement the remaining entries should turn out to be zero since they give the covariance between amplitude and phase quadratures. The small deviations from zero show that the measurements were not perfectly orthogonal but close.
To verify that our source is indeed an Einstein-Podolsky-Rosen source we calculated the EPR-Reid covariance product~\cite{Reid1989}
\begin{equation*}
    \min_g\Var(X_A-gX_B) \cdot \min_h\Var(P_A-hP_B) < 1\ ,
\end{equation*}
which was $0.31 < 1$ for our states setting a new benchmark with entanglement generated by a squeezed mode superimposed with a vacuum mode.

Using the recorded data we analysed the feasibility to use our state for quantum key distribution. For all calculations we assumed the covariance matrix of Eq.~(\ref{eqn:covariance matrix}) to be the reconstructed covariance matrix in the parameter estimation phase of the quantum key distribution protocol, regardless for how many samples the key rate was calculated. We then constructed the confidence set assuming $k=10^6$ samples were used to estimate the correlation terms between Alice and Bob in the covariance matrix. All diagonal terms were assumed to be estimated by using the total of $N$ samples measured in one quadrature by each party. 

Figure~\ref{fig:keyrate_vs_samples_X} shows the calculated key rate for our state versus the total number of measured samples when omitting all samples measured in the $P$ quadrature (anti-squeezed quadrature), i.e.\ $p_X = 1$. The key rate is given in secure bits per sample that can be used to generate the raw key, i.e.\ the samples left after sifting, parameter estimation and omission of samples measured in the $P$ quadrature. The parameters used for the calculation are $\epsilon_{\text{pe}} = \epsilon_c = \epsilon_s = 10^{-16}$ and $\beta=0.9$ for the efficiency of the error correction. $\alpha$ is chosen $8$ times the standard deviation of the quadrature given by the covariance matrix of Eq.~(\ref{eqn:covariance matrix}). Each curve in the figure was calculated for a different number of bins which was taken as $2^{n_{\text{bits}}}$ by choosing an apropriate $\delta$. For $n_{\text{bits}} \ge 6$ the asymptotic maximum of the key rate is reached. For $10^9$ samples, which is experimentally challenging but achievable~\cite{Jouguet2012}, the key rate is already close to the maximum value of about $0.7$\,bits per sample and even for $10^8$ samples it is not much lower.

Figure~\ref{fig:keyrate_vs_samples_P} shows the same as Fig.~\ref{fig:keyrate_vs_samples_X} but with the samples from the $X$ quadrature omitted, i.e.\ $p_X=0$. To reach the asymptotic value for the key rate $n_{\text{bits}} \ge 7$ is needed. In comparison to the key rate for only the $X$ quadrature samples the asymptotic key rate is lower and a larger number of samples is necessary for a positive key rate. Here also about $10^9$ samples are necessary to reach a value close to the maximum and also for $10^8$ samples the key rate is not much lower.

In Figure~\ref{fig:keyrate_vs_samples_XP} samples from both $X$ and $P$ quadrature measurements were used to compute the secure key rate, i.e.\ $p_X=0.5$. For the calculation we used $n_{\text{bits}} = 6$ for the $X$ quadrature and $n_{\text{bits}} = 7$ for the $P$ quadrature. Although the number of secure bits per sample is smaller than for considering only samples from the $X$ quadrature, the number of samples used to generate the raw key is almost twice as large leading to a larger number of secret keys for the same number of measured samples.

\begin{figure}[t]
    \includegraphics[width=13cm]{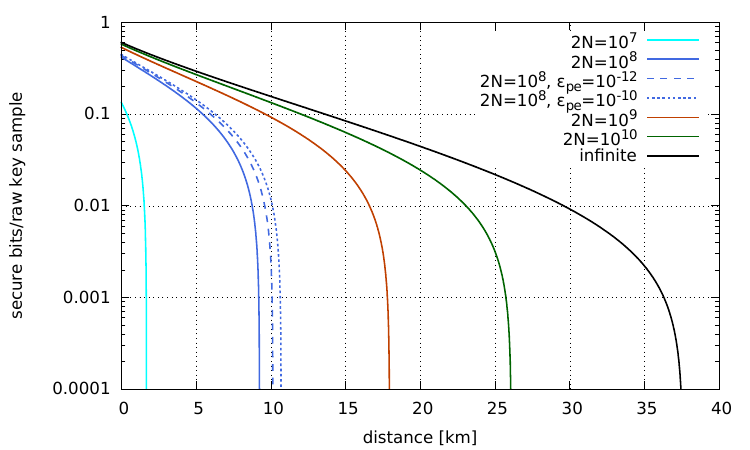}
    \caption{Key rate versus distance when sending one part of the entangled beam through an optical fiber. The key rate is given as the number of secure bits per sample that can be used to generate a key, i.e.\ after sifting and parameter estimation. We assumed a coupling efficiency of $95$\,\% into the optical fiber and an optical loss of $0.2$\,dB/km. The curves are plotted for different total number of measured samples $2N$. The parameter estimation security parameter was chosen $\epsilon_{\text{pe}} = 10^{-16}$ except for the dashed lines.}
    \label{fig:keyrate_vs_distance}
\end{figure}

When sending Bob's part of the entangled state through an optical fiber the key rate versus the distance is shown in Fig.~\ref{fig:keyrate_vs_distance}. We assumed a coupling efficiency of $95$\,\% into the optical fiber~\cite{Mehmet2010} and an optical loss of $0.2$\,dB/km. To maximize the achievable distance we omitted the samples from the $P$ quadrature. Considering only samples from the $P$ quadrature would yield a maximal distance of only about $2$\,km and considering both $X$ and $P$ quadrature measurements would yield about $8$\,km for an infinite number of samples. The parameters used for the calculation were chosen as for Fig.~\ref{fig:keyrate_vs_samples}, in particular $n_{\text{bits}} = 6$ and $\epsilon_{\text{pe}}=10^{-16}$. From the figure we read that the absolute maximal distance for our state is about $37$\,km. Taking a realistic number of samples~\cite{Jouguet2012} the reachable distance shrinks to about $9$\,km for $10^8$ samples and about $18$\,km for $10^9$ samples. These values are limited by the parameter estimation. Curves with a relaxed parameter estimation security parameter of $\epsilon_{\text{pe}} = 10^{-12}$ and $\epsilon_{\text{pe}} = 10^{-10}$ are shown for $10^8$ samples in the figure. For $\epsilon_{\text{pe}} = 10^{-10}$~\cite{Leverrier2010} the distance increases to about $10.5$\,km. Increasing the number of samples $k$ used for parameter estimation instead does not change the achievable distances as $\text{Cov}(X_A,X_B)$ is not critical for our states and $\text{Cov}(P_A,P_B)$ can be estimated more precisely as all measurements where Alice and Bob measured the $P$ quadrature simultaneously, can be used. Here $\text{Cov}$ denotes the covariance. For the calculation we assumed that no excess noise is introduced by the fiber as stated in~\cite{Lodewyck2005}. Excess noise introduced by the electronic dark noise of the homodyne detectors is instead already included in the reconstructed covariance matrix. As the error correction efficiency depends on the given signal-to-noise ratio, the actually maximal achievable distance depends on the availability of efficient error correcting codes~\cite{Jouguet2012}.

\section{Conclusion}
\label{sec: conclusion}
We have presented an analysis of a gaussian entanglement source involving a squeezed mode and a vacuum mode regarding entanglement-based quantum key distribution under the assumption of collective attacks including finite-size effects. While in the present experiment the entanglement has been distributed on an optical table, coupling one part of the bipartite state into a standard optical telecommunication fiber and building Bob's detector remotely would allow for quantum key distribution in local-area networks. The local oscillator for Bob's homodyne detector could be served e.g.\ from an auxiliary laser at Bob's site which could be phase locked to the control beam accompanying the entangled state. This scheme also ensures that phase noise introduced by the fiber is not crucial below the unity-gain frequency of the phase-locked loop. Our analysis revealed that a distance of more than ten kilometers is possible with a reasonable but challenging number of measured samples of $10^9$ even though a vacuum mode was included in the generation of the states. As for achieving a large distance only samples measured in the squeezed quadrature are used, a probability larger than $50$\,\% for measuring the squeezed quadrature compared to the anti-squeezed quadrature would increase the number of secure bits generated per second. Since a squeezing bandwidth of more than $100$\,MHz was already demonstrated~\cite{Mehmet2010a}, a QKD system involving a single squeezed-light source could achieve significant overall key rates.

Although the restriction to a single squeezed input mode, as presented here, reduces the complexity of the source, a full scheme with two squeezed fields superimposed at a balanced beam splitter will achieve higher key rates. For the full scheme the achievable distance for $10^9$ samples would be about $28$\,km in comparison to about $17$\,km for a single squeezed-light source. While these values were calculated for an entangled state generated by two identical squeezed modes, our security analysis provided here also allows the use of states generated by squeezed modes with different squeezing strengths.

\ack
This research was supported by the EU FP\,7 project Q-ESSENCE (Grant agreement number 248095). TE and VH thank the IMPRS on Gravitational Wave Astronomy for support. VH acknowledges support from HALOSTAR. TF acknowledges support from DFG under grant WE-1240/12-1 and from BMBF project QUOREP. FF acknowledges support from Japan Society for the Promotion of Science (JSPS) by KAKENHI grants No. 24-02793. RFW acknowledges support by the EU FP\,7 project COQUIT (Grant agreement number 233747).

\end{document}